\documentclass[amsmath,amssymb,amsfonts,aps,prl,twocolumn,showpacs,superscriptaddress]{revtex4-1}

\usepackage{graphicx}
\usepackage{dcolumn}
\usepackage{bm}
\usepackage{color}
\usepackage[normalem]{ulem}

\begin{document}

\newcommand{\bra}[1]{\left\langle#1\right|}
\newcommand{\ket}[1]{\left|#1\right\rangle}
\newcommand{\bracket}[2]{\big\langle#1 \bigm| #2\big\rangle}

\newcommand{\Tr}{{\rm Tr}}
\renewcommand{\Im}{{\rm Im}}
\renewcommand{\Re}{{\rm Re}}

\newcommand{\ra}{\rightarrow}

\newcommand{\xc}{{\rm xc}}
\newcommand{\Hxc}{{\rm Hxc}}
\newcommand{\SSM}{{\rm SSM}}
\newcommand{\imp}{{\rm imp}}
\newcommand{\loc}{{\rm loc}}

\newcommand{\su}{\uparrow}
\newcommand{\sd}{\downarrow}

\newcommand{\tip}{{\rm tip}}
\newcommand{\It}{\tilde{I}}

\newcommand{\david}[1]{{\color{red}#1}}
\newcommand{\stkout}[1]{\ifmmode\text{\sout{\ensuremath{#1}}}\else\sout{#1}\fi}

\newcommand{\be}{\begin{equation}}
\newcommand{\ee}{\end{equation}}
\newcommand{\bea}{\begin{eqnarray}}
\newcommand{\eea}{\end{eqnarray}}
\newcommand\diag[2]{%
  \ncline[linewidth=1pt,nodesep=-30pt]{#1}{#2}}

\def\a{\alpha}
\def\b{\beta}
\def\g{\gamma}
\def\G{\Gamma}
\def\d{\delta}
\def\D{\Delta}
\def\e{\epsilon}
\def\ve{\varepsilon}
\def\h{\eta}
\def\th{\theta}
\def\k{\kappa}
\def\l{\lambda}
\def\L{\Lambda}
\def\m{\mu}
\def\n{\nu}
\def\c{\xi}
\def\C{\Xi}
\def\p{\pi}
\def\P{\Pi}
\def\r{\rho}
\def\s{\sigma}
\def\S{\Sigma}
\def\t{\tau}
\def\f{\phi}
\def\vf{\varphi}
\def\F{\Phi}
\def\x{\chi}
\def\w{\omega}
\def\W{\Omega}
\def\q{\psi}
\def\Q{\Psi}
\def\z{\zeta}

\def\bga{\mbox{\boldmath $\alpha$}}
\def\bgb{\mbox{\boldmath $\beta$}}
\def\bgg{\mbox{\boldmath $\gamma$}}
\def\bgG{\mbox{\boldmath $\Gamma$}}
\def\bgd{\mbox{\boldmath $\delta$}}
\def\bgD{\mbox{\boldmath $\Delta$}}
\def\bge{\mbox{\boldmath $\epsilon$}}
\def\bgve{\mbox{\boldmath $\varepsilon$}}
\def\bgh{\mbox{\boldmath $\eta$}}
\def\bgth{\mbox{\boldmath $\theta$}}
\def\bgk{\mbox{\boldmath $\kappa$}}
\def\bgl{\mbox{\boldmath $\lambda$}}
\def\bgL{\mbox{\boldmath $\Lambda$}}
\def\bgm{\mbox{\boldmath $\mu$}}
\def\bgn{\mbox{\boldmath $\nu$}}
\def\bgc{\mbox{\boldmath $\xi$}}
\def\bgC{\mbox{\boldmath $\Xi$}}
\def\bgp{\mbox{\boldmath $\pi$}}
\def\bgP{\mbox{\boldmath $\Pi$}}
\def\bgr{\mbox{\boldmath $\rho$}}
\def\bgs{\mbox{\boldmath $\sigma$}}
\def\bgS{\mbox{\boldmath $\Sigma$}}
\def\bgt{\mbox{\boldmath $\tau$}}
\def\bgf{\mbox{\boldmath $\phi$}}
\def\bgvf{\mbox{\boldmath $\varphi$}}
\def\bgF{\mbox{\boldmath $\Phi$}}
\def\bgx{\mbox{\boldmath $\chi$}}
\def\bgw{\mbox{\boldmath $\omega$}}
\def\bgW{\mbox{\boldmath $\Omega$}}
\def\bgq{\mbox{\boldmath $\psi$}}
\def\bgQ{\mbox{\boldmath $\Psi$}}
\def\bgz{\mbox{\boldmath $\zeta$}}

\def\calgG{\mbox{$\mathit{\Gamma}$}}  
\def\calgD{\mbox{$\mathit{\Delta}$}}  
\def\calgL{\mbox{$\mathit{\Lambda}$}}  
\def\calgC{\mbox{$\mathit{\Xi}$}}  
\def\calgP{\mbox{$\mathit{\Pi}$}}  
\def\calgS{\mbox{$\mathit{\Sigma}$}} 
\def\calgF{\mbox{$\mathit{\Phi}$}} 
\def\calgW{\mbox{$\mathit{\Omega}$}} 
\def\calgQ{\mbox{$\mathit{\Psi}$}} 

\def\bcalgG{\mbox{\boldmath $\mathit{\Gamma}$}}  
\def\bcalgD{\mbox{\boldmath $\mathit{\Delta}$}}  
\def\bcalgL{\mbox{\boldmath $\mathit{\Lambda}$}}  
\def\bcalgC{\mbox{\boldmath $\mathit{\Xi}$}}  
\def\bcalgP{\mbox{\boldmath $\mathit{\Pi}$}}  
\def\bcalgS{\mbox{\boldmath $\mathit{\Sigma}$}} 
\def\bcalgF{\mbox{\boldmath $\mathit{\Phi}$}} 
\def\bcalgW{\mbox{\boldmath $\mathit{\Omega}$}} 
\def\bcalgQ{\mbox{\boldmath $\mathit{\Psi}$}} 

\def\bla{{\mathbf a}}
\def\blb{{\mathbf b}}
\def\blc{{\mathbf c}}
\def\bld{{\mathbf d}}
\def\ble{{\mathbf e}}
\def\blf{{\mathbf f}}
\def\blg{{\mathbf g}}
\def\blh{{\mathbf h}}
\def\bli{{\mathbf i}}
\def\blj{{\mathbf j}}
\def\blk{{\mathbf k}}
\def\bll{{\mathbf l}}
\def\blm{{\mathbf m}}
\def\bln{{\mathbf n}}
\def\blo{{\mathbf o}}
\def\blp{{\mathbf p}}
\def\blq{{\mathbf q}}
\def\blr{{\mathbf r}}
\def\bls{{\mathbf s}}
\def\blt{{\mathbf t}}
\def\blu{{\mathbf u}}
\def\blv{{\mathbf v}}
\def\blw{{\mathbf w}}
\def\blx{{\mathbf x}}
\def\bly{{\mathbf y}}
\def\blz{{\mathbf z}}

\def\bcalla{\mbox{\boldmath $a$}}
\def\bcallb{\mbox{\boldmath $b$}}
\def\bcallc{\mbox{\boldmath $c$}}
\def\bcalld{\mbox{\boldmath $d$}}
\def\bcalle{\mbox{\boldmath $e$}}
\def\bcallf{\mbox{\boldmath $f$}}
\def\bcallg{\mbox{\boldmath $g$}}
\def\bcallh{\mbox{\boldmath $h$}}
\def\bcalli{\mbox{\boldmath $i$}}
\def\bcallj{\mbox{\boldmath $j$}}
\def\bcallk{\mbox{\boldmath $k$}}
\def\bcalll{\mbox{\boldmath $l$}}
\def\bcallm{\mbox{\boldmath $m$}}
\def\bcalln{\mbox{\boldmath $n$}}
\def\bcallo{\mbox{\boldmath $o$}}
\def\bcallp{\mbox{\boldmath $p$}}
\def\bcallq{\mbox{\boldmath $q$}}
\def\bcallr{\mbox{\boldmath $r$}}
\def\bcalls{\mbox{\boldmath $s$}}
\def\bcallt{\mbox{\boldmath $t$}}
\def\bcallu{\mbox{\boldmath $u$}}
\def\bcallv{\mbox{\boldmath $v$}}
\def\bcallw{\mbox{\boldmath $w$}}
\def\bcallx{\mbox{\boldmath $x$}}
\def\bcally{\mbox{\boldmath $y$}}
\def\bcallz{\mbox{\boldmath $z$}}

\def\blA{{\mathbf A}}
\def\blB{{\mathbf B}}
\def\blC{{\mathbf C}}
\def\blD{{\mathbf D}}
\def\blE{{\mathbf E}}
\def\blF{{\mathbf F}}
\def\blG{{\mathbf G}}
\def\blH{{\mathbf H}}
\def\blI{{\mathbf I}}
\def\blJ{{\mathbf J}}
\def\blK{{\mathbf K}}
\def\blL{{\mathbf L}}
\def\blM{{\mathbf M}}
\def\blN{{\mathbf N}}
\def\blO{{\mathbf O}}
\def\blP{{\mathbf P}}
\def\blQ{{\mathbf Q}}
\def\blR{{\mathbf R}}
\def\blS{{\mathbf S}}
\def\blT{{\mathbf T}}
\def\blU{{\mathbf U}}
\def\blV{{\mathbf V}}
\def\blW{{\mathbf W}}
\def\blX{{\mathbf X}}
\def\blY{{\mathbf Y}}
\def\blZ{{\mathbf Z}}

\def\bilA{\mbox{\boldmath $A$}}
\def\bilB{\mbox{\boldmath $B$}}
\def\bilC{\mbox{\boldmath $C$}}
\def\bilD{\mbox{\boldmath $D$}}
\def\bilE{\mbox{\boldmath $E$}}
\def\bilF{\mbox{\boldmath $F$}}
\def\bilG{\mbox{\boldmath $G$}}
\def\bilH{\mbox{\boldmath $H$}}
\def\bilI{\mbox{\boldmath $I$}}
\def\bilJ{\mbox{\boldmath $J$}}
\def\bilK{\mbox{\boldmath $K$}}
\def\bilL{\mbox{\boldmath $L$}}
\def\bilM{\mbox{\boldmath $M$}}
\def\bilN{\mbox{\boldmath $N$}}
\def\bilO{\mbox{\boldmath $O$}}
\def\bilP{\mbox{\boldmath $P$}}
\def\bilQ{\mbox{\boldmath $Q$}}
\def\bilR{\mbox{\boldmath $R$}}
\def\bilS{\mbox{\boldmath $S$}}
\def\bilT{\mbox{\boldmath $T$}}
\def\bilU{\mbox{\boldmath $U$}}
\def\bilV{\mbox{\boldmath $V$}}
\def\bilW{\mbox{\boldmath $W$}}
\def\bilX{\mbox{\boldmath $X$}}
\def\bilY{\mbox{\boldmath $Y$}}
\def\bilZ{\mbox{\boldmath $Z$}}

\def\callA{\mbox{$\mathcal{A}$}}
\def\callB{\mbox{$\mathcal{B}$}}
\def\callC{\mbox{$\mathcal{C}$}}
\def\callD{\mbox{$\mathcal{D}$}}
\def\callE{\mbox{$\mathcal{E}$}}
\def\callF{\mbox{$\mathcal{F}$}}
\def\callG{\mbox{$\mathcal{G}$}}
\def\callH{\mbox{$\mathcal{H}$}}
\def\callI{\mbox{$\mathcal{I}$}}
\def\callJ{\mbox{$\mathcal{J}$}}
\def\callK{\mbox{$\mathcal{K}$}}
\def\callL{\mbox{$\mathcal{L}$}}
\def\callM{\mbox{$\mathcal{M}$}}
\def\callN{\mbox{$\mathcal{N}$}}
\def\callO{\mbox{$\mathcal{O}$}}
\def\callP{\mbox{$\mathcal{P}$}}
\def\callQ{\mbox{$\mathcal{Q}$}}
\def\callR{\mbox{$\mathcal{R}$}}
\def\callS{\mbox{$\mathcal{S}$}}
\def\callT{\mbox{$\mathcal{T}$}}
\def\callU{\mbox{$\mathcal{U}$}}
\def\callV{\mbox{$\mathcal{V}$}}
\def\callW{\mbox{$\mathcal{W}$}}
\def\callX{\mbox{$\mathcal{X}$}}
\def\callY{\mbox{$\mathcal{Y}$}}
\def\callZ{\mbox{$\mathcal{Z}$}}

\def\bcallA{\mbox{\boldmath $\mathcal{A}$}}
\def\bcallB{\mbox{\boldmath $\mathcal{B}$}}
\def\bcallC{\mbox{\boldmath $\mathcal{C}$}}
\def\bcallD{\mbox{\boldmath $\mathcal{D}$}}
\def\bcallE{\mbox{\boldmath $\mathcal{E}$}}
\def\bcallF{\mbox{\boldmath $\mathcal{F}$}}
\def\bcallG{\mbox{\boldmath $\mathcal{G}$}}
\def\bcallH{\mbox{\boldmath $\mathcal{H}$}}
\def\bcallI{\mbox{\boldmath $\mathcal{I}$}}
\def\bcallJ{\mbox{\boldmath $\mathcal{J}$}}
\def\bcallK{\mbox{\boldmath $\mathcal{K}$}}
\def\bcallL{\mbox{\boldmath $\mathcal{L}$}}
\def\bcallM{\mbox{\boldmath $\mathcal{M}$}}
\def\bcallN{\mbox{\boldmath $\mathcal{N}$}}
\def\bcallO{\mbox{\boldmath $\mathcal{O}$}}
\def\bcallP{\mbox{\boldmath $\mathcal{P}$}}
\def\bcallQ{\mbox{\boldmath $\mathcal{Q}$}}
\def\bcallR{\mbox{\boldmath $\mathcal{R}$}}
\def\bcallS{\mbox{\boldmath $\mathcal{S}$}}
\def\bcallT{\mbox{\boldmath $\mathcal{T}$}}
\def\bcallU{\mbox{\boldmath $\mathcal{U}$}}
\def\bcallV{\mbox{\boldmath $\mathcal{V}$}}
\def\bcallW{\mbox{\boldmath $\mathcal{W}$}}
\def\bcallX{\mbox{\boldmath $\mathcal{X}$}}
\def\bcallY{\mbox{\boldmath $\mathcal{Y}$}}
\def\bcallZ{\mbox{\boldmath $\mathcal{Z}$}}


\def\ua{\uparrow}
\def\da{\downarrow}
\def\ra{\rightarrow}
\def\la{\leftarrow}
\def\La{\Leftarrow}
\def\Ra{\Rightarrow}
\def\de{\partial}
\def\iif{\infty}
\def\bra{\langle}
\def\ket{\rangle}
\def\grad{\mbox{\boldmath $\nabla$}}
\def\Tr{{\rm Tr}}
\def\Re{{\rm Re}}
\def\Im{{\rm Im}}


\def\iu{{\rm i}}
\def\1op{\hat{\mathbbm{1}}}
\def\nn{\nonumber}
\def\AA{\mathring{\mathrm{A}}}

\title{Mott metal-insulator transition from steady-state density functional theory}

\author{David Jacob}
\affiliation{Nano-Bio Spectroscopy Group and European Theoretical Spectroscopy
  Facility (ETSF), Dpto. Pol\'{i}meros y Materiales Avanzados: F\'{i}sica,
  Qu\'{i}mica y Tecnolog\'{i}a, Universidad del Pa\'{i}s Vasco UPV/EHU,
  Av. Tolosa 72, E-20018 San Sebasti\'{a}n, Spain}
\affiliation{IKERBASQUE, Basque Foundation for Science, Plaza Euskadi 5, E-48009 Bilbao, Spain}

\author{Gianluca Stefanucci}
\affiliation{Dipartimento di Fisica, Universit\`{a} di Roma Tor Vergata,
Via della Ricerca Scientifica 1, 00133 Rome, Italy}
\affiliation{INFN, Sezione di Roma Tor Vergata, Via della Ricerca Scientifica 1, 00133 Rome, Italy}

\author{Stefan Kurth}
\affiliation{Nano-Bio Spectroscopy Group and European Theoretical Spectroscopy
  Facility (ETSF), Dpto. Pol\'{i}meros y Materiales Avanzados: F\'{i}sica,
  Qu\'{i}mica y Tecnolog\'{i}a, Universidad del Pa\'{i}s Vasco UPV/EHU,
  Av. Tolosa 72, E-20018 San Sebasti\'{a}n, Spain}
\affiliation{IKERBASQUE, Basque Foundation for Science, Plaza Euskadi 5, E-48009 Bilbao, Spain}
\affiliation{Donostia International Physics Center (DIPC), Paseo Manuel de
  Lardizabal 4, E-20018 San Sebasti\'{a}n, Spain}

\begin{abstract}
We present a computationally efficient method to obtain the spectral 
function of bulk systems in the framework of steady-state density functional
theory (i-DFT) using an idealized Scanning Tunneling Microscope (STM) setup.
We calculate the current through the STM tip and 
then extract the spectral function from the finite-bias differential 
conductance. The fictitious non-interacting system of i-DFT features 
an exchange-correlation (xc) contribution to the bias which 
guarantees the same current as in the true interacting system. 
Exact properties of the xc bias are established using Fermi-liquid 
theory and subsequently implemented to construct approximations for 
the Hubbard model. We show for two different lattice structures 
that the Mott metal-insulator transition is captured by i-DFT.
\end{abstract}

\date{\today}

\maketitle

{\em Introduction.--}
Standard wisdom has it that density functional theory (DFT)
\cite{DreizlerGross:90} is not capable of 
describing strongly correlated materials. 
The origin of this misconception is twofold. By construction, the exact
exchange-correlation (xc) potential of the Kohn-Sham (KS) system yields 
the exact electronic density and
directly related quantities. However, approximations to the 
xc potential often miss the step features due to the derivative discontinuity
of the  generating xc energy functional~\cite{PerdewParrLevyBalduz:82}. These
features are a crucial ingredient to capture strong correlation effects in
diverse physical situations such as, e.g., molecular dissociation
\cite{RuzsinszkyPerdewCsonkaVydrovScuseria:06,FuksMaitra:14},
fermion gases in optical lattices
\cite{Xianlong_etal:06}
or transport
\cite{StefanucciKurth:11,BergfieldLiuBurkeStafford:12,TroesterSchmitteckertEvers:12,KurthStefanucci:13,KurthStefanucci:17,DittmannSplettstoesserHelbig:18,DittmannHelbigKennes:19}.
Approximations which include the steps are under active development
\cite{MirtschinkSeidlGoriGiorgi:13,KraislerKronik:13,BroscoYingLorenzana:13,YingBroscoLorenzana:14,Sobrino:PRB:2020}. 
Furthermore, the interpretation of the KS excitation 
energies as true excitation energies is not rigorously justified, 
even if the exact xc potential is used. 
While in the limit of weak
correlations this may be a reasonable approximation, it completely fails 
in the opposite limit -- it is easy to show that the exact KS band structure
of the Hubbard model in the Mott insulating phase has no gap and the
derivative discontinuity plays a crucial role in describing the Mott transition
\cite{LimaSilvaOliveiraCapelle:03,KarlssonPriviteiraVerdozzi:11,KVOC.2011,KKPV.2013}. 

In general, extracting excitation energies in a DFT framework is not 
straightforward. While charge neutral excitations are accessible via
time-dependent (TD) DFT \cite{RungeGross:84,Ullrich:12}, excitations which do
change the number of electrons such as those probed in (inverse) photoemission
are encoded in the spectral function, an arduous quantity to calculate also
for TDDFT~\cite{Uimonen_2014}.
Usually spectral functions are calculated within a Green's function 
framework~\cite{MartinReiningCeperley:16,svl-book}, e.g. GW~\cite{Aryasetiawan_1998,GolzeDvorakRinke_2019}, 
Dynamical Mean-Field Theory (DMFT)~\cite{Metzner:PRL:1989,Georges:PRB:1992,Georges:RMP:1996} 
and GW+DMFT~\cite{Biermann:PRL:2003,Biermann_2014}, 
but these methods come at considerable computational cost.
Instead, DMFT combined with DFT offers a pragmatic approach to compute the spectra of strongly correlated 
materials~\cite{Lichtenstein_1998,Vollhardt_2005,Kotliar:RMP:2006}, although the double counting problem remains unsolved.

Recently we proposed a method to compute 
the spectral function~\cite{JacobKurth:18}
of a nanoscale tunneling 
junction using an extension of DFT, called steady-state DFT or i-DFT
\cite{StefanucciKurth:15}.
In i-DFT the fundamental variables are
the non-equilibrium steady-state
density of and current through the junction. 
Hence the KS system
requires  a nonequilibrium extension of the standard xc potential as
well as the introduction of an xc contribution to the applied bias in 
the electrodes~\cite{sa-1.2004,sa-2.2004,SaiZwolakVignaleDiVentra:05,KoentoppBurkeEvers:06}. 
In an idealized Scanning Tunneling Microscope (STM) setup where one of the 
electrodes (i.e. the ``STM tip'') couples only weakly to the 
nanoscale junction, the 
spectral function at frequency $\w$
can be obtained from the differential
conductance at bias $V=\w$~\cite{JacobKurth:18,KurthJacobSobrinoStefanucci:19}.

In this Letter we generalize the i-DFT+STM approach to calculate the 
spectral function of
arbitrary bulk systems. We further show that 
the Mott metal-insulator (MI) transition in the Hubbard 
model, one of the main paradigms in the field of strongly correlated
electrons, can be described by i-DFT provided that both the xc potential and  
the xc bias feature steps as function of the steady density {\em and} current.
General properties of the xc 
bias are derived using Fermi liquid (FL) theory in combination with DMFT~\cite{Metzner:PRL:1989,Georges:PRB:1992,Georges:RMP:1996}.
Taking advantage of ideas developed previously in the context of the Anderson impurity
model \cite{StefanucciKurth:15,KurthStefanucci:16} we construct an approximation satisfying all 
FL+DMFT properties and illustrate the MI transition in two different crystal structures, 
the Bethe and the cubic lattices.

{\em Bulk spectral function from i-DFT.--}
We consider a bulk system described by a Hamiltonian written in terms of 
creation and annihilation operators for electrons with spin 
projection $\s$ in basis functions $\{\vf_{i}\}$. The basis functions are taken 
orthonormal but otherwise completely general -- they can be, e.g., 
extended Bloch states or localized Wannier functions.
No assumptions on the explicit form of the Hamiltonian is made at this stage. 
The system is probed by an ideal nonmagnetic tip, a fictitious
``Gedanken'' device with the following properties: (i) the electrons in the
tip are noninteracting with energy dispersion $\e_{k}$ and wavefunctions
$\q_{k}$. This property ensures the applicability of the Meir-Wingreen
formula \cite{MeirWingreen:92} for the steady current from the tip to the
bulk \cite{JacobKurth:18}; (ii) the coupling of the tip to the bulk is weak
but otherwise its form can be chosen freely. For convenience, here we take it
to be coupled exclusively to the 
$n$-th basis function $\vf_{n}$ of the bulk. Letting $T_{k}$ be the 
one-electron integral between the states  $\q_{k}$ and $\vf_{n}$, 
the ideal tip is chosen to have a transition rate
$\g=2\p\sum_{k}|T_{k}|^{2}\d(\w-\e_{k})$ 
independent of $\w$ (wide band limit). 
Without any loss of generality we set the chemical potential of the 
whole system (tip plus bulk) to zero.
A bias $V$ is applied only in the tip and as a consequence 
a steady current $I(V)$ flows toward the bulk through state $\vf_{n}$. 
In the limit of vansihing coupling, the bulk remains
in equilibrium and its spectral function
projected onto the state $\vf_{n}$ can then be written as~\cite{JacobKurth:18}
\be
A(\w)=\lim_{\g\to 0}\frac{\p}{\g}\left.\frac{dI(V)}{dV}\right|_{V=\w}.
\label{AfromI}
\ee

Here we use i-DFT to compute $I(V)$. 
In i-DFT the bulk density for a given potential $v$ and the
steady current for a given bias $V$  are reproduced in
the same but {\em noninteracting} bulk system coupled to
the same tip. This fictitious KS system is subject to the effective
potential $v_s = v +  v_{\rm Hxc}$, where $v_{\rm Hxc}$ is Hartree
plus exchange-correlation (Hxc) potential, and to the 
effective bias $V_s = V+V_{\rm xc}$ with $V_{\rm xc}$ the xc bias.
Both $v_{\rm Hxc}$ and
$V_{\rm xc}$ are functionals of the bulk density and the steady current.
However, in the $\g\to 0$ limit, see Eq.~(\ref{AfromI}), $v_{\rm Hxc}$ becomes
independent of $I$ and approaches the ground-state Hxc potential
of DFT~\cite{JacobKurth:18}. In what 
follows we assume that $v_{\rm Hxc}$ is known from a 
previous DFT calculation. Denoting by $A_{s}(\w)$ the 
ground-state KS spectral function we then have
\be
\lim_{\g\to 0}\frac{I}{\g}=\int\frac{d\w}{\p}[f(\w-V-V_{\rm 
xc})-f(\w)]A_{s}(\w).
\label{IfromAs}
\ee
For any given bias $V$ this equation must be solved self-consistently 
since $V_{\rm xc}$ depends on $I$. The relation between  
$A$ and  $A_{s}$ follows directly from the 
derivative of Eq.~(\ref{IfromAs}) with respect to $V$
\begin{align}
A(\w)=\lim_{\g\to 0}\frac{A_{s}(\w+V_{\rm xc})}
{1-\frac{\g}{\p}\frac{dV_{\rm xc}}{dI}A_{s}(\w+V_{\rm xc})}
=A_{s}(\W)\frac{d\W}{d\w}
\label{JKformula}
\end{align}
where $\W(\w)=\w+V_{\rm xc}[I(\w)]$. Equation~(\ref{JKformula})
is one of the main results of this Letter and shows that
i-DFT can be used to calculate bulk spectral functions.

{\em Properties of the xc bias from Fermi-liquid theory.--}
For i-DFT to become a practical and computationally efficient 
scheme we need to develop accurate approximations to $V_{\rm xc}$.
Any approximation should satisfy $V_{\rm xc}[0]=0$, for 
otherwise there would be a finite current at zero bias.
Below we derive a few more properties for uniform systems
from FL theory and DMFT. 
We concentrate on the local description (hence $\vf_{n}$ is a site 
basis function) and use DMFT which becomes exact in the limit of infinite dimensions 
(or, more rigorously, coordination number)
\cite{Metzner:PRL:1989,Georges:PRB:1992,Georges:RMP:1996} -- 
and otherwise yields a very good approximation for dimensions $\ge3$~\cite{Florens:PRB:2004}.

Due to the Friedel sum rule~\cite{mera-2,Schmitteckert:PRL:2008,TroesterSchmitteckertEvers:12}, 
the spectral function evaluated at the Fermi energy depends only
on the bulk density. As the latter is the same in the many-body and the KS
system we have 
$A(0)=A_{s}(0)$. Since $I(0)=0$ then $\W(0)=V_{\rm xc}[0]=0$ and therefore 
Eq.~(\ref{JKformula}) implies
\be
\left.\frac{d V_{\rm xc}}{dI}\right|_{I=0}=0.
\label{d1vxc}
\ee

Other properties can be obtained for particle-hole (ph) symmetric  
systems, e.g., the half-filled Hubbard model on bipartite lattices. 
In DMFT the local Green's function can be written as
$G^{-1}(\w)=\w-v-\tilde{\S}(\w)-\D_{0}(\w)$ where $v$ is the uniform 
potential, $\D_{0}(\w)=\L_{0}(\w)-i\G_{0}(\w)/2$ the {\em noninteracting} 
embedding self-energy (or hybridization function) and $\tilde{\S}(\w)$ 
the many-body self-energy.
We emphasize that $\tilde{\S}(\w)$ is not the local DMFT 
self-energy $\Sigma(\w)$ as it contains also correlation corrections to the 
embedding:
\be
\tilde{\S}=\S+\D-\D_{0}.
\label{stildes}
\ee
At half-filling the spectral function $A(\w)=i[G(\w)-G^{\ast}(\w)]$
is an even function of frequency. Additionally,
the ph symmetry yields a condition for the Hxc potential, i.e.,
$v_{\rm Hxc}=-v$. Hence the KS Green's function is simply 
$G_{s}(\w)=\left[\w-\D_{0}(\w)\right]^{-1}$ and
therefore the KS spectral function 
$A_{s}(\w)=i[G_{s}(\w)-G_{s}^{\ast}(\w)]$ is even too.

Differentiating Eq.~(\ref{JKformula}), evaluating the result in 
$\w=0$ and using $A'(0)=A'_{s}(0)=0$ (henceforth primes are used to 
denote derivatives with respect to $\w$) we find $A_{s}(0)\W''(0)=0$.
Using $A_{s}(0)=A(0)=\lim_{\g\to 0}\frac{\p}{\g}I'(0)\neq 0$, 
see Eq.~(\ref{AfromI}), then yields
\be
\W''(0)=\left.\frac{d^{2}V_{\rm 
xc}}{dI^{2}}\right|_{I=0}I'(0)^{2}+
\left.\frac{d V_{\rm 
xc}}{dI}\right|_{I=0}I''(0)=0.
\ee
Taking further into account Eq.~(\ref{d1vxc}), this implies that also 
the second derivative of $V_{\rm xc}$ w.r.t. the current should vanish:
\be
\left.\frac{d^{2}V_{\rm 
xc}}{dI^{2}}\right|_{I=0}=0.
\label{d2vxc}
\ee
The third derivative of $V_{\rm xc}$ w.r.t. the current
is nonvanishing and can be related to the pseudo quasi-particle weight
\be
\tilde{Z}^{-1}\equiv 
1-{\rm Re}[\tilde{\S}'(0)].
\ee
In the Supplemental Material~\footnote{
See the Supplemental Material at XXX for the detailed proofs of Eqs.~(\ref{d3vxc}) and (\ref{ztildez}), as well
as details for the reverse-engineering and parametrization of the xc-bias functional.}
we prove that
\be
\left.\frac{d^{3}V_{\rm xc}}{dI^{3}}\right|_{I=0}=
-\frac{\p^{3}\G_{0}(0)}{8\g^{3}}\left[(\tilde{Z}^{-1}-\L'_{0}(0))^{2}-
(1-\L'_{0}(0))^{2}\right].
\label{d3vxc}
\ee
We shall use the properties in 
Eqs.~(\ref{d1vxc}), (\ref{d2vxc}) and (\ref{d3vxc}) to construct  
approximations to the xc bias.

We observe that the pseudo quasi-particle weight can be expressed in 
terms of the 
actual quasi-particle weight $Z\equiv [1-{\rm Re}[\S'(0)]^{-1}$ through
Eq.~(\ref{stildes}):
$\tilde{Z}^{-1}=Z^{-1}-\Re[\D'(0)]+\Re[\D_{0}'(0)]$. In DMFT 
the interacting embedding self-energy (or hybridization function) $\D(\w)$ is related to 
the local Green's function $G(\w)=N^{-1}\sum_k(\w-\e_k-\S(\w))^{-1}$ via the Dyson equation
$\D(\w)=\w-v-\S-[G(\w)]^{-1}$. Differentiation and evaluation at $\w=0$ then yields
$\D^\prime(0)=Z^{-1}+G^\prime(0)/[G(0)]^2$. While $G(0)=G_s(0)$ by Friedel sum rule, 
it is straightforward to show \cite{Note1} that 
$G^\prime(0)=Z^{-1}G_s^\prime(0)$.  Thus
\be
\tilde{Z}^{-1} = 1 + \frac{Z-1}{Z} \cdot \frac{G_s^\prime(0)}{[G_s(0)]^2}
\label{ztildez}
\ee
which can be easily evaluated since $G_s(\omega)$ 
depends only on the lattice properties through $\D_{0}(\w)$.

{\em The xc bias for the metal-insulator transition.--}
We consider
the half-filled Hubbard model on the Bethe lattice (BL) with infinite coordination 
number as well as 
on a cubic lattice (CL), and describe the strategy common to the construction of
the xc bias for both lattices. 

In the insulating Mott phase the Hubbard bands become
Coulomb blockade (CB) peaks as the hopping integral between 
neighboring sites decreases. In the limit of vanishing hopping the CB peaks are 
separated in energy by the Hubbard interaction $U$. 
For finite hopping the CB peaks are
separated by the discontinuity $U_{\rm xc}$ of the Hxc potential
(with $\lim_{U\to \infty} U_{\rm xc}(U)/U = 1$). The xc bias should feature a step
of height $U_{\rm xc}$ \cite{JacobKurth:18,KurthJacob:18}, 
\be
\bar{V}_{\rm xc}[I]=-\frac{U_{\rm xc}}{2} {\rm sign}(\tilde{I})-
(U-U_{\rm xc}) g(\tilde{I}) \;.
\label{barvxc}
\ee
Here we have defined the reduced current as $\tilde{I}=I/(2\g)$ and its 
physically realizable domain is $\tilde{I}\in(-\frac{1}{2},\frac{1}{2})$.
The  function $g(\tilde{I})$ depends on the lattice and
fulfills the general properties: (i) 
$g(-\tilde{I})=-g(\tilde{I})$, and
(ii) 
$g(\pm\frac{1}{2})= \pm \frac{1}{2}$ such that
$\bar{V}_{\rm xc}[\pm \g] = \mp \frac{U}{2}$. 
In the SI we describe the strategy to obtain 
$U_{\rm xc}(U)$ and  $g(\tilde{I})$ for both the Bethe and the cubic 
lattice. 

The approximation in Eq.~(\ref{barvxc}) violates the property in 
Eq.~(\ref{d1vxc}) which  is crucial for
describing the Kondo peak in the metallic 
phase~\cite{KurthStefanucci:16}. We then make the ansatz
\be
V_{\rm xc}[I]=a(I)\bar{V}_{\rm xc}[I].
\label{xcbias}
\ee
For $I$ close to zero we can approximate $\bar{V}_{\rm xc}[I]\simeq 
-\frac{U_{\rm xc}}{2}{\rm sign}(\tilde{I})$. If $\tilde{I}$ is nonnegative 
we can rewrite this expression as the limiting function of the 
sequence $\bar{V}^{(n)}_{\rm xc}[I]=-\frac{U_{\rm xc}}{2}\tilde{I}^{1/n}$.
Letting $\a\tilde{I}^{p}$ be the leading order term of the function 
$a(I)$ as $\tilde{I}\to 0^{+}$ we have $V^{(n)}_{\rm xc}[I]\simeq -\frac{\a 
U_{\rm xc}}{2}\tilde{I}^{p+1/n}$. We then see
that the properties in 
Eqs.~(\ref{d1vxc}), (\ref{d2vxc}) and (\ref{d3vxc})  are 
fulfilled provided that $p=3+1/n$. Taking the limit $n\to\iif$ and 
using that $a(I)=a(-I)$ we 
infer that for $I\simeq 0$ the function $a(I)\simeq 
\a|\tilde{I}|^{3}$. In the following we parametrize this function as
\be
a(I)=\frac{2}{\p}{\rm atan}\left(|\l_{\rm K}\tilde{I}|^{3}\right)
\label{a(I)}
\ee
since for $|\tilde{I}|=1/2$ we must have $a(I)\simeq 1$ for 
 $V_{\rm xc}[\pm \g] \simeq \mp 
\frac{U}{2}$. Taking into account that 
\be
\left.\frac{d^{3}V_{\rm xc}}{dI^{3}}\right|_{I=0}=-\frac{3U_{\rm 
xc}}{4\p\g^{3}}\l_{\rm K}^{3},
\label{d3vxc2}
\ee
the 
parameter $\l_{\rm K}\gg 1$ can be determined from
Eq.~(\ref{d3vxc}).

\begin{figure}[t]
  \includegraphics[width=\linewidth]{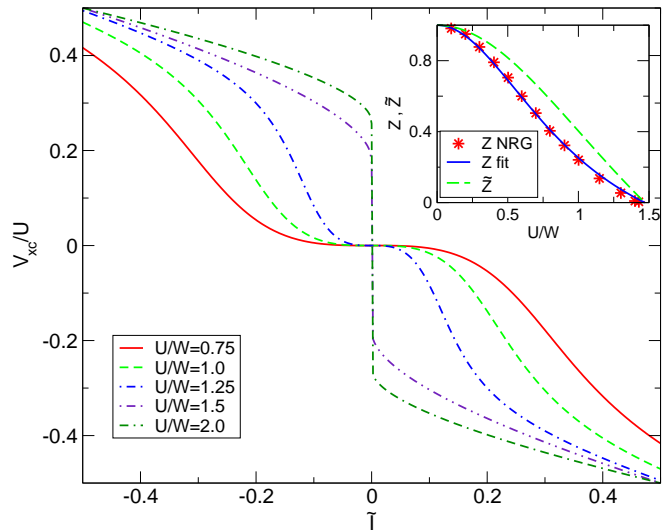}
  \caption{Model xc bias of Eq.~(\ref{xcbias}) for the Bethe lattice for
    different values of $U$. Inset: Quasiparticle weights $Z$ from NRG of
    Ref.~\cite{Bulla:99}, our fit to the NRG data as well as $\tilde{Z}$
    according to Eq.~(\ref{ztildez}). $W$ is the width of the band
    for $U=0$.}
\label{bethe_quantities}
\end{figure}

We apply the i-DFT approach to the calculation of the spectral
function of the Hubbard model on a BL and CL.
To obtain $\bar{V}_{\rm xc}[I]$ in Eq.~(\ref{barvxc})
we performed DMFT calculations for both lattices using the 
non-crossing approximation (NCA)~\cite{PruschkeCoxJarrell:93} in the
insulating phase. Reverse engineering the DMFT+NCA spectral 
function \cite{Note1} we found that an accurate parametrization is provided by
\be
g(\tilde{I})=\left((1-b)\sqrt{|\tilde{I}/2|}+b\tilde{I}\right){\rm 
sign}(\tilde{I}),
\ee 
with $b=1/4$ in the BL and $b=0$ in the CL, whereas $U_{\rm xc}(U)$ is a 
smooth increasing function of $U$, see Ref.~\cite{Note1} for the explicit form.
To determine $\l_{\rm K}$ and hence the function $a(I)$ we first consider
a BL. In this case 
$\D_{0}(\w)=\frac{\w}{2}-i\sqrt{\left(\frac{W}{4}\right)^{2}-\left(\frac{\w}{2}\right)^{2}}$, where $W$ is the bandwidth; 
hence $\L'(0)=1/2$ and $\G_{0}(0)=W/2$. Inserting these values in Eq.~(\ref{d3vxc}) 
and using Eq.~(\ref{d3vxc2}) we obtain
\be
\l_{\rm K}^{3}=\frac{\p^{4}W}{12 U_{\rm xc}}\frac{1-\tilde{Z}}{\tilde{Z}^{2}}. 
\ee
Close to zero frequency the BL Green's function 
$G(\w)=(4/W)^2 \D(\w) \simeq Z/(\w-Z\D(\w))$, which implies 
$\D(\w)=\D_{0}(\w/Z)$; hence from
Eq.~(\ref{ztildez}) 
\be
\tilde{Z}=\frac{2Z}{1+Z}.
\ee
The quasi-particle weight has been accurately calculated in
Ref.~\onlinecite{Bulla:99} using NRG, and it is well approximated by a
shifted Lorentzian~\cite{Note1}. In the 
inset of Fig.~\ref{bethe_quantities} we show the NRG $Z$, our fit 
and the pseudo quasi-particle weight $\tilde{Z}$. Proceeding along 
the same lines we constructed the xc bias for a CL, see Ref.~\cite{Note1} for 
details. We anticipate that  $\tilde{Z}(Z)$ is almost identical in 
the two lattices.

\begin{figure}[t]
\includegraphics[width=\linewidth]{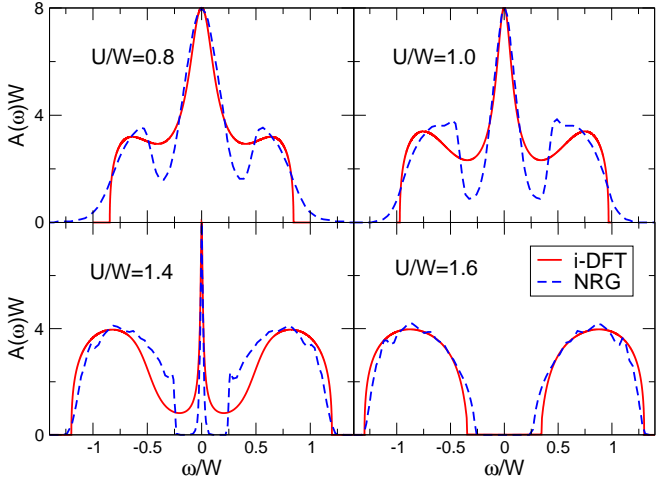}
\caption{Spectral functions of the Hubbard model on the Bethe lattice for
  different interaction strengths obtained by i-DFT and compared with the NRG
  results of Ref.~\cite{ZitkoPruschke:09}. $W$ is the width of the band
  for $U=0$.}
\label{bethespectra}
\end{figure}

{\em Results.--}
In Fig.~\ref{bethe_quantities} we show the BL xc bias
for different values of $U$ in units of the non-interacting bandwidth $W$. 
In the metallic phase, $U/W<1.3$, $V_{\rm xc}$ exhibits a plateau around
$\tilde{I}= 0$ which turns into a sharp step
in the insulating phase. The development of a step is essential for 
the gap opening, see below.

In Fig.~\ref{bethespectra} we compare the i-DFT spectral functions
with NRG results from Ref.~\onlinecite{ZitkoPruschke:09} for different
interaction strengths.  i-DFT captures the essential features
of the spectra such as the Kondo peak at $\w=0$ in the metallic phase as
well as its disappearance with increasing interaction strength. The curvature
of the Kondo peak at $\w=0$ is, by construction, correctly described by our
xc bias but also the Hubbard side bands are captured reasonably well,
especially for large $U$'s. The approximation to $V_{\rm xc}$ performs poorly 
in the frequency range of the minima of $A(\w)$ 
and some of the finer features of the NRG spectra
are also missing. Interestingly, however, the i-DFT
spectra always have finite support. This can be understood from
Eq.~(\ref{JKformula}) by noting that (i) $A_s(\w)$ has finite support and
(ii) the xc bias is restricted to the intervall
$V_{\rm xc}\in[-\frac{U}{2},\frac{U}{2}]$. 

\begin{figure}[t]
\includegraphics[width=\linewidth]{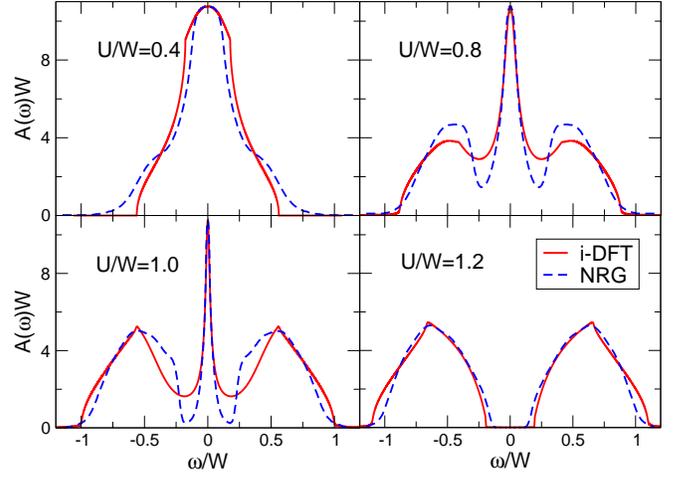}
\caption{Spectral functions of the Hubbard model on the simple cubic lattice
  for different interaction strengths obtained by i-DFT and compared with the
  NRG results of Ref.~\cite{ZitkoBoncaPruschke:09}. }
\label{cubicspectra}
\end{figure}

The performance of the approximate xc bias for a CL is illustrated in 
Fig.~\ref{cubicspectra} where we again compare i-DFT and NRG~\cite{ZitkoBoncaPruschke:09} 
spectral functions for different interaction strengths. 
The general trend is similar to the
previous case, in particular the MI transition is
correctly captured. One feature which draws attention
is the presence of ``kinks'' in the i-DFT spectra which are directly
attributable to the van Hove singularities in the KS density of states.

{\em Conclusions.--}
In summary, we have shown how one can extract bulk spectral functions from
i-DFT. A particular emphasis has been on the proper description of the Mott 
metal-insulator transition in strongly correlated systems which so far has
been elusive within DFT. We have derived properties of the
crucial i-DFT quantity, the xc bias, by establishing a connection to
Fermi-liquid theory. These properties, together with DMFT and NRG
reference calculations, have been employed
to construct approximations for the Hubbard model on the
infinitely coordinated Bethe lattice as well as on the cubic lattice.
The approximated xc bias differs from previous ones used in the Anderson 
model, which is always metallic due to the Kondo peak at the Fermi energy.

For any given lattice the i-DFT potentials $v_{\rm Hxc}$ and $V_{\rm 
xc}$ are ``universal'', 
i.e., they are independent of the external on-site potential and 
bias (these are the conjugate variables  to the density and current,
respectively). 
Therefore, the potentials derived here can also serve to calculate the 
spectral function of Hubbard systems with,  e.g., nonmagnetic impurities or 
disorder, in the same spirit as in DFT an accurate parametrization of 
$v_{\rm Hxc}$ for the homogeneous electron gas is used, through the 
local density approximation, to deal with inhomogeneous systems. 

Although the xc bias is lattice dependent, our work 
highlights important general features, namely the step of height 
$U_{\rm xc}$ and the dependence of the Kondo prefactor on 
the pesudo quasi-particle weight. As the underlying i-DFT theorem 
makes no assumption on dimensionality, these features provide  
important guidance for the design of i-DFT functionals in lower 
dimensions.

Last but not least, the i-DFT spectra capture the essential physics of the
Mott metal-insulator transition at negligible computational cost,
paving the way to an ab-initio description of strongly
correlated solids within a density functional framework.

{\em Acknowledgments.--} D.J. and S.K. acknowledge funding through
a grant ``Grupos Consolidados UPV/EHU del Gobierno Vasco''
(Grant No. IT1249-19). G.S. acknowledges funding from
MIUR PRIN Grant No. 20173B72NB and from INFN17-Nemesys project.

\bibliographystyle{prstyown}
\bibliography{bib_spectra}

\end{document}


\newcommand{\bra}[1]{\left\langle#1\right|}
\newcommand{\ket}[1]{\left|#1\right\rangle}
\newcommand{\bracket}[2]{\big\langle#1 \bigm| #2\big\rangle}

\newcommand{\Tr}{{\rm Tr}}
\renewcommand{\Im}{{\rm Im}}
\renewcommand{\Re}{{\rm Re}}

\newcommand{\ra}{\rightarrow}

\newcommand{\xc}{{\rm xc}}
\newcommand{\Hxc}{{\rm Hxc}}
\newcommand{\SSM}{{\rm SSM}}
\newcommand{\imp}{{\rm imp}}
\newcommand{\loc}{{\rm loc}}

\newcommand{\su}{\uparrow}
\newcommand{\sd}{\downarrow}

\newcommand{\tip}{{\rm tip}}
\newcommand{\It}{\tilde{I}}

\newcommand{\be}{\begin{equation}}
\newcommand{\ee}{\end{equation}}
\newcommand{\bea}{\begin{eqnarray}}
\newcommand{\eea}{\end{eqnarray}}
\newcommand\diag[2]{%
  \ncline[linewidth=1pt,nodesep=-30pt]{#1}{#2}}

\def\a{\alpha}
\def\b{\beta}
\def\g{\gamma}
\def\G{\Gamma}
\def\d{\delta}
\def\D{\Delta}
\def\e{\epsilon}
\def\ve{\varepsilon}
\def\h{\eta}
\def\th{\theta}
\def\k{\kappa}
\def\l{\lambda}
\def\L{\Lambda}
\def\m{\mu}
\def\n{\nu}
\def\c{\xi}
\def\C{\Xi}
\def\p{\pi}
\def\P{\Pi}
\def\r{\rho}
\def\s{\sigma}
\def\S{\Sigma}
\def\t{\tau}
\def\f{\phi}
\def\vf{\varphi}
\def\F{\Phi}
\def\x{\chi}
\def\w{\omega}
\def\W{\Omega}
\def\q{\psi}
\def\Q{\Psi}
\def\z{\zeta}

\def\bga{\mbox{\boldmath $\alpha$}}
\def\bgb{\mbox{\boldmath $\beta$}}
\def\bgg{\mbox{\boldmath $\gamma$}}
\def\bgG{\mbox{\boldmath $\Gamma$}}
\def\bgd{\mbox{\boldmath $\delta$}}
\def\bgD{\mbox{\boldmath $\Delta$}}
\def\bge{\mbox{\boldmath $\epsilon$}}
\def\bgve{\mbox{\boldmath $\varepsilon$}}
\def\bgh{\mbox{\boldmath $\eta$}}
\def\bgth{\mbox{\boldmath $\theta$}}
\def\bgk{\mbox{\boldmath $\kappa$}}
\def\bgl{\mbox{\boldmath $\lambda$}}
\def\bgL{\mbox{\boldmath $\Lambda$}}
\def\bgm{\mbox{\boldmath $\mu$}}
\def\bgn{\mbox{\boldmath $\nu$}}
\def\bgc{\mbox{\boldmath $\xi$}}
\def\bgC{\mbox{\boldmath $\Xi$}}
\def\bgp{\mbox{\boldmath $\pi$}}
\def\bgP{\mbox{\boldmath $\Pi$}}
\def\bgr{\mbox{\boldmath $\rho$}}
\def\bgs{\mbox{\boldmath $\sigma$}}
\def\bgS{\mbox{\boldmath $\Sigma$}}
\def\bgt{\mbox{\boldmath $\tau$}}
\def\bgf{\mbox{\boldmath $\phi$}}
\def\bgvf{\mbox{\boldmath $\varphi$}}
\def\bgF{\mbox{\boldmath $\Phi$}}
\def\bgx{\mbox{\boldmath $\chi$}}
\def\bgw{\mbox{\boldmath $\omega$}}
\def\bgW{\mbox{\boldmath $\Omega$}}
\def\bgq{\mbox{\boldmath $\psi$}}
\def\bgQ{\mbox{\boldmath $\Psi$}}
\def\bgz{\mbox{\boldmath $\zeta$}}

\def\calgG{\mbox{$\mathit{\Gamma}$}}  
\def\calgD{\mbox{$\mathit{\Delta}$}}  
\def\calgL{\mbox{$\mathit{\Lambda}$}}  
\def\calgC{\mbox{$\mathit{\Xi}$}}  
\def\calgP{\mbox{$\mathit{\Pi}$}}  
\def\calgS{\mbox{$\mathit{\Sigma}$}} 
\def\calgF{\mbox{$\mathit{\Phi}$}} 
\def\calgW{\mbox{$\mathit{\Omega}$}} 
\def\calgQ{\mbox{$\mathit{\Psi}$}} 

\def\bcalgG{\mbox{\boldmath $\mathit{\Gamma}$}}  
\def\bcalgD{\mbox{\boldmath $\mathit{\Delta}$}}  
\def\bcalgL{\mbox{\boldmath $\mathit{\Lambda}$}}  
\def\bcalgC{\mbox{\boldmath $\mathit{\Xi}$}}  
\def\bcalgP{\mbox{\boldmath $\mathit{\Pi}$}}  
\def\bcalgS{\mbox{\boldmath $\mathit{\Sigma}$}} 
\def\bcalgF{\mbox{\boldmath $\mathit{\Phi}$}} 
\def\bcalgW{\mbox{\boldmath $\mathit{\Omega}$}} 
\def\bcalgQ{\mbox{\boldmath $\mathit{\Psi}$}} 

\def\bla{{\mathbf a}}
\def\blb{{\mathbf b}}
\def\blc{{\mathbf c}}
\def\bld{{\mathbf d}}
\def\ble{{\mathbf e}}
\def\blf{{\mathbf f}}
\def\blg{{\mathbf g}}
\def\blh{{\mathbf h}}
\def\bli{{\mathbf i}}
\def\blj{{\mathbf j}}
\def\blk{{\mathbf k}}
\def\bll{{\mathbf l}}
\def\blm{{\mathbf m}}
\def\bln{{\mathbf n}}
\def\blo{{\mathbf o}}
\def\blp{{\mathbf p}}
\def\blq{{\mathbf q}}
\def\blr{{\mathbf r}}
\def\bls{{\mathbf s}}
\def\blt{{\mathbf t}}
\def\blu{{\mathbf u}}
\def\blv{{\mathbf v}}
\def\blw{{\mathbf w}}
\def\blx{{\mathbf x}}
\def\bly{{\mathbf y}}
\def\blz{{\mathbf z}}

\def\bcalla{\mbox{\boldmath $a$}}
\def\bcallb{\mbox{\boldmath $b$}}
\def\bcallc{\mbox{\boldmath $c$}}
\def\bcalld{\mbox{\boldmath $d$}}
\def\bcalle{\mbox{\boldmath $e$}}
\def\bcallf{\mbox{\boldmath $f$}}
\def\bcallg{\mbox{\boldmath $g$}}
\def\bcallh{\mbox{\boldmath $h$}}
\def\bcalli{\mbox{\boldmath $i$}}
\def\bcallj{\mbox{\boldmath $j$}}
\def\bcallk{\mbox{\boldmath $k$}}
\def\bcalll{\mbox{\boldmath $l$}}
\def\bcallm{\mbox{\boldmath $m$}}
\def\bcalln{\mbox{\boldmath $n$}}
\def\bcallo{\mbox{\boldmath $o$}}
\def\bcallp{\mbox{\boldmath $p$}}
\def\bcallq{\mbox{\boldmath $q$}}
\def\bcallr{\mbox{\boldmath $r$}}
\def\bcalls{\mbox{\boldmath $s$}}
\def\bcallt{\mbox{\boldmath $t$}}
\def\bcallu{\mbox{\boldmath $u$}}
\def\bcallv{\mbox{\boldmath $v$}}
\def\bcallw{\mbox{\boldmath $w$}}
\def\bcallx{\mbox{\boldmath $x$}}
\def\bcally{\mbox{\boldmath $y$}}
\def\bcallz{\mbox{\boldmath $z$}}

\def\blA{{\mathbf A}}
\def\blB{{\mathbf B}}
\def\blC{{\mathbf C}}
\def\blD{{\mathbf D}}
\def\blE{{\mathbf E}}
\def\blF{{\mathbf F}}
\def\blG{{\mathbf G}}
\def\blH{{\mathbf H}}
\def\blI{{\mathbf I}}
\def\blJ{{\mathbf J}}
\def\blK{{\mathbf K}}
\def\blL{{\mathbf L}}
\def\blM{{\mathbf M}}
\def\blN{{\mathbf N}}
\def\blO{{\mathbf O}}
\def\blP{{\mathbf P}}
\def\blQ{{\mathbf Q}}
\def\blR{{\mathbf R}}
\def\blS{{\mathbf S}}
\def\blT{{\mathbf T}}
\def\blU{{\mathbf U}}
\def\blV{{\mathbf V}}
\def\blW{{\mathbf W}}
\def\blX{{\mathbf X}}
\def\blY{{\mathbf Y}}
\def\blZ{{\mathbf Z}}

\def\bilA{\mbox{\boldmath $A$}}
\def\bilB{\mbox{\boldmath $B$}}
\def\bilC{\mbox{\boldmath $C$}}
\def\bilD{\mbox{\boldmath $D$}}
\def\bilE{\mbox{\boldmath $E$}}
\def\bilF{\mbox{\boldmath $F$}}
\def\bilG{\mbox{\boldmath $G$}}
\def\bilH{\mbox{\boldmath $H$}}
\def\bilI{\mbox{\boldmath $I$}}
\def\bilJ{\mbox{\boldmath $J$}}
\def\bilK{\mbox{\boldmath $K$}}
\def\bilL{\mbox{\boldmath $L$}}
\def\bilM{\mbox{\boldmath $M$}}
\def\bilN{\mbox{\boldmath $N$}}
\def\bilO{\mbox{\boldmath $O$}}
\def\bilP{\mbox{\boldmath $P$}}
\def\bilQ{\mbox{\boldmath $Q$}}
\def\bilR{\mbox{\boldmath $R$}}
\def\bilS{\mbox{\boldmath $S$}}
\def\bilT{\mbox{\boldmath $T$}}
\def\bilU{\mbox{\boldmath $U$}}
\def\bilV{\mbox{\boldmath $V$}}
\def\bilW{\mbox{\boldmath $W$}}
\def\bilX{\mbox{\boldmath $X$}}
\def\bilY{\mbox{\boldmath $Y$}}
\def\bilZ{\mbox{\boldmath $Z$}}

\def\callA{\mbox{$\mathcal{A}$}}
\def\callB{\mbox{$\mathcal{B}$}}
\def\callC{\mbox{$\mathcal{C}$}}
\def\callD{\mbox{$\mathcal{D}$}}
\def\callE{\mbox{$\mathcal{E}$}}
\def\callF{\mbox{$\mathcal{F}$}}
\def\callG{\mbox{$\mathcal{G}$}}
\def\callH{\mbox{$\mathcal{H}$}}
\def\callI{\mbox{$\mathcal{I}$}}
\def\callJ{\mbox{$\mathcal{J}$}}
\def\callK{\mbox{$\mathcal{K}$}}
\def\callL{\mbox{$\mathcal{L}$}}
\def\callM{\mbox{$\mathcal{M}$}}
\def\callN{\mbox{$\mathcal{N}$}}
\def\callO{\mbox{$\mathcal{O}$}}
\def\callP{\mbox{$\mathcal{P}$}}
\def\callQ{\mbox{$\mathcal{Q}$}}
\def\callR{\mbox{$\mathcal{R}$}}
\def\callS{\mbox{$\mathcal{S}$}}
\def\callT{\mbox{$\mathcal{T}$}}
\def\callU{\mbox{$\mathcal{U}$}}
\def\callV{\mbox{$\mathcal{V}$}}
\def\callW{\mbox{$\mathcal{W}$}}
\def\callX{\mbox{$\mathcal{X}$}}
\def\callY{\mbox{$\mathcal{Y}$}}
\def\callZ{\mbox{$\mathcal{Z}$}}

\def\bcallA{\mbox{\boldmath $\mathcal{A}$}}
\def\bcallB{\mbox{\boldmath $\mathcal{B}$}}
\def\bcallC{\mbox{\boldmath $\mathcal{C}$}}
\def\bcallD{\mbox{\boldmath $\mathcal{D}$}}
\def\bcallE{\mbox{\boldmath $\mathcal{E}$}}
\def\bcallF{\mbox{\boldmath $\mathcal{F}$}}
\def\bcallG{\mbox{\boldmath $\mathcal{G}$}}
\def\bcallH{\mbox{\boldmath $\mathcal{H}$}}
\def\bcallI{\mbox{\boldmath $\mathcal{I}$}}
\def\bcallJ{\mbox{\boldmath $\mathcal{J}$}}
\def\bcallK{\mbox{\boldmath $\mathcal{K}$}}
\def\bcallL{\mbox{\boldmath $\mathcal{L}$}}
\def\bcallM{\mbox{\boldmath $\mathcal{M}$}}
\def\bcallN{\mbox{\boldmath $\mathcal{N}$}}
\def\bcallO{\mbox{\boldmath $\mathcal{O}$}}
\def\bcallP{\mbox{\boldmath $\mathcal{P}$}}
\def\bcallQ{\mbox{\boldmath $\mathcal{Q}$}}
\def\bcallR{\mbox{\boldmath $\mathcal{R}$}}
\def\bcallS{\mbox{\boldmath $\mathcal{S}$}}
\def\bcallT{\mbox{\boldmath $\mathcal{T}$}}
\def\bcallU{\mbox{\boldmath $\mathcal{U}$}}
\def\bcallV{\mbox{\boldmath $\mathcal{V}$}}
\def\bcallW{\mbox{\boldmath $\mathcal{W}$}}
\def\bcallX{\mbox{\boldmath $\mathcal{X}$}}
\def\bcallY{\mbox{\boldmath $\mathcal{Y}$}}
\def\bcallZ{\mbox{\boldmath $\mathcal{Z}$}}


\def\ua{\uparrow}
\def\da{\downarrow}
\def\ra{\rightarrow}
\def\la{\leftarrow}
\def\La{\Leftarrow}
\def\Ra{\Rightarrow}
\def\de{\partial}
\def\iif{\infty}
\def\bra{\langle}
\def\ket{\rangle}
\def\grad{\mbox{\boldmath $\nabla$}}
\def\Tr{{\rm Tr}}
\def\Re{{\rm Re}}
\def\Im{{\rm Im}}


\def\iu{{\rm i}}
\def\1op{\hat{\mathbbm{1}}}
\def\nn{\nonumber}
\def\AA{\mathring{\mathrm{A}}}

\renewcommand\theequation{S\arabic{equation}}
\renewcommand\thefigure{S\arabic{figure}}

\title{Supplemental Material: The Mott metal-insulator transition from steady-state density functional theory}

\author{David Jacob}
\affiliation{Nano-Bio Spectroscopy Group and European Theoretical Spectroscopy
Facility (ETSF), Dpto. de F\'{i}sica de Materiales,
Universidad del Pa\'{i}s Vasco UPV/EHU, Av. Tolosa 72,
E-20018 San Sebasti\'{a}n, Spain}
\affiliation{IKERBASQUE, Basque Foundation for Science, Maria Diaz de Haro 3,
E-48013 Bilbao, Spain}

\author{Gianluca Stefanucci}
\affiliation{Dipartimento di Fisica, Universit\`{a} di Roma Tor Vergata,
Via della Ricerca Scientifica 1, 00133 Rome, Italy; European Theoretical
Spectroscopy Facility (ETSF)}
\affiliation{INFN, Laboratori Nazionali di Frascati, Via E. Fermi 40,
00044 Frascati, Italy}

\author{Stefan Kurth}
\affiliation{Nano-Bio Spectroscopy Group and European Theoretical Spectroscopy
Facility (ETSF), Dpto. de F\'{i}sica de Materiales,
Universidad del Pa\'{i}s Vasco UPV/EHU, Av. Tolosa 72,
E-20018 San Sebasti\'{a}n, Spain}
\affiliation{IKERBASQUE, Basque Foundation for Science, Maria Diaz de Haro 3,
E-48013 Bilbao, Spain}
\affiliation{Donostia International Physics Center (DIPC), Paseo Manuel de
  Lardizabal 4, E-20018 San Sebasti\'{a}n, Spain}

\maketitle

\section{Proof of Eq.~(9) in main text}

Eq.~(9) in the main text relates the third derivative of the xc-bias w.r.t. the current, 
$\partial^3_I{V}_\xc(0)$, to the pseudo quasiparticle weight $\tilde{Z}$ defined in Eq.~(8) 
in the main text. 
The starting point for the proofs is Eq.~(3) in the main text which relates
the true many-body spectral function $A(\w)$ to the KS spectral function 
$A_s(\w)$ which can be written in a very compact form as:
\be
A(\w)= A_{s}(\W) \, \frac{d\W}{d\w}  \hspace{1ex} \mbox{ where } \hspace{1ex} \W(\w) \equiv \w + V_\xc[I(\w)].
\label{JKSformula}
\ee
Differentiating $A(\w)$ (denoted by primes as in the main text) twice then yields:
\bea
A^\prime(\w) &=& A^\prime_s(\W) \, (\W^\prime(\w))^2 + A_s(\W) \, \W^{\prime\prime}(\w)
\label{A1}
\\
A^{\prime\prime}(\w) &=& A_s^{\prime\prime}(\W) \, (\W^\prime(\w))^3 + 
3 \, A_s^\prime(\W) \, \W^\prime(\w) \, \W^{\prime\prime}(\w) +
A_s(\W) \, \W^{\prime\prime\prime}(\w).
\label{A2}
\eea
We thus need to know up to the third derivative of $\W(\w)$ w.r.t. $\omega$: 
\bea
\W^\prime(\w) &=& 1 + \frac{\partial{I}}{\partial\w} \, \frac{\partial{V}_\xc}{\partial{I}}
= 1 + \frac{\gamma}{\pi} \, A(\w) \, \frac{\partial{V}_\xc}{\partial{I}}
\label{Om1}
\\
\W^{\prime\prime}(\w) &=& \frac{\gamma}{\pi} \, A^\prime(\w) \, \frac{\partial{V}_\xc}{\partial{I}}
+ \frac{\gamma}{\pi} \, A(\w) \, \frac{\partial{I}}{\partial\w} \, \frac{\partial^2{V}_\xc}{\partial{I}^2}
=  \frac{\gamma}{\pi} \, A^\prime(\w) \, \frac{\partial{V}_\xc}{\partial{I}}
+ \left(\frac{\gamma}{\pi}\right)^2 \, (A(\w))^2 \, \frac{\partial^2{V}_\xc}{\partial{I}^2}
\label{Om2}
\\
\W^{\prime\prime\prime}(\w) &=& 
\frac{\gamma}{\pi} \, A^{\prime\prime}(\w) \, \frac{\partial{V}_\xc}{\partial{I}} + 
3 \left(\frac{\gamma}{\pi}\right)^2 \, A^\prime(\w) \, A(\w) \, \frac{\partial^2{V}_\xc}{\partial{I}^2} + 
\left(\frac{\gamma}{\pi}\right)^3 \, (A(\w))^3 \, \frac{\partial^3{V}_\xc}{\partial{I}^3} 
\label{Om3}
\eea
where we have made use of $\partial{I}/\partial\w=\frac{\gamma}{\pi}A(\w)$ (Eq.~(1) in the main text).
Evaluating (\ref{A2}) at $\w=0$ and taking into account that $\partial_I{V}_\xc|_{I=0}=\partial^2_I{V}_\xc|_{I=0}=0$
(Eqs.~(4) and (7) in main text) so that $\W^\prime(0)=1$ and $\W^{\prime\prime}(0)=0$, we find
\be
A^{\prime\prime}(0) = A_s^{\prime\prime}(0) + A_s(0) \, \W^{\prime\prime\prime}(0)
= A_s^{\prime\prime}(0) + \left(\frac{\gamma}{\pi}\right)^3 \, (A_s(0))^4 \, 
\left.\frac{\partial^3{V}_\xc}{\partial{I}^3}\right|_{I=0}.
\label{d2A}
\ee
On the other hand, close to zero frequency the local GF $G(\w)$ can be written in terms of the pseudo
quasiparticle weight $\tilde{Z}$ defined by Eq.~(8) in the main text as $G(\w)\simeq\tilde{Z}/(\w-\tilde{Z}\,\D_{0}(\w))$.
The corresponding spectral function $A(\w)=i[G(\w)-G^\ast(\w)]$ can then be written as
\be
A(\w)\simeq 
\tilde{Z}^{2}\frac{\G_{0}(\w)}{(\w-\tilde{Z}\L_{0}(\w))^{2}+\tilde{Z}^{2}\G_{0}(\w)^{2}/4}.
\label{Aapprox}
\ee
Differentiating $A(\w)$ twice and evaluating at $\w=0$ yields after some simple but lengthy algebra:
\be
A''(0)=-4\frac{\G_{0}''(0)}{\G_{0}(0)^{2}}-32
\frac{(\tilde{Z}^{-1}-\L'_{0}(0))^{2}}{\G_{0}(0)^{3}}.
\label{d2Aqp}
\ee
Combining (\ref{d2A}) and (\ref{d2Aqp}) and solving for $\left.\frac{\partial^3{V}_\xc}{\partial{I}^3}\right|_{I=0}$ then yields
Eq.~(9) in the main text.

\section{Relation between $G^\prime(0)$ and $G_s^\prime(0)$}

For the proof of Eq.~(10) in the main text relating the pseudo quasiparticle weight $\tilde{Z}$ and the actual quasiparticle
weight $Z$, we make use of a simple relationship between the derivative of the local interacting GF $G^\prime(0)$ and the derivative of the non-interacting (i.e. Kohn-Sham) GF $G_s^\prime(0)$ within DMFT. The proof is straightforward.
The starting point is the representation of the local GF in k-space within DMFT as
\be
G(\w) = \frac{1}{N} \sum_k \frac{1}{\w-\e_k-\S(\w)} 
\label{Gloc}
\ee
where $\e_k$ is the band dispersion, $\S(\w)$ is the local self-energy, and $N$ is the number of sampling points in k-space.
The derivative of the GF is given by:
\be
G^\prime(\w) = -\frac{1}{N} \sum_k \frac{1-\Sigma^\prime(\w)}{(\w-\e_k-\S(\w))^2}.
\label{Gloc1}
\ee
At half filling, evaluation at $\w=0$ then yields:
\be
G^\prime(0) = -\frac{1}{N} \sum_k \frac{Z^{-1}}{(\e_k)^2} = Z^{-1} \, G_s^\prime(0)
\label{Gloc1_0}
\ee
where we have used the definition of the quasiparticle weight as $Z=(1-\S^\prime(0))^{-1}$ and the fact that $G_s^\prime(0)=-N^{-1}\sum_k1/(\e_k)^2$ which follows simply by setting $\S(\w)=0$ in Eqs. (\ref{Gloc}-\ref{Gloc1_0}) above.

\section{Details of the parametrization of the xc-bias functional}

\subsection{Reverse-engineering of xc-bias from DMFT+NCA spectra}

\begin{figure}
  \includegraphics[width=\linewidth]{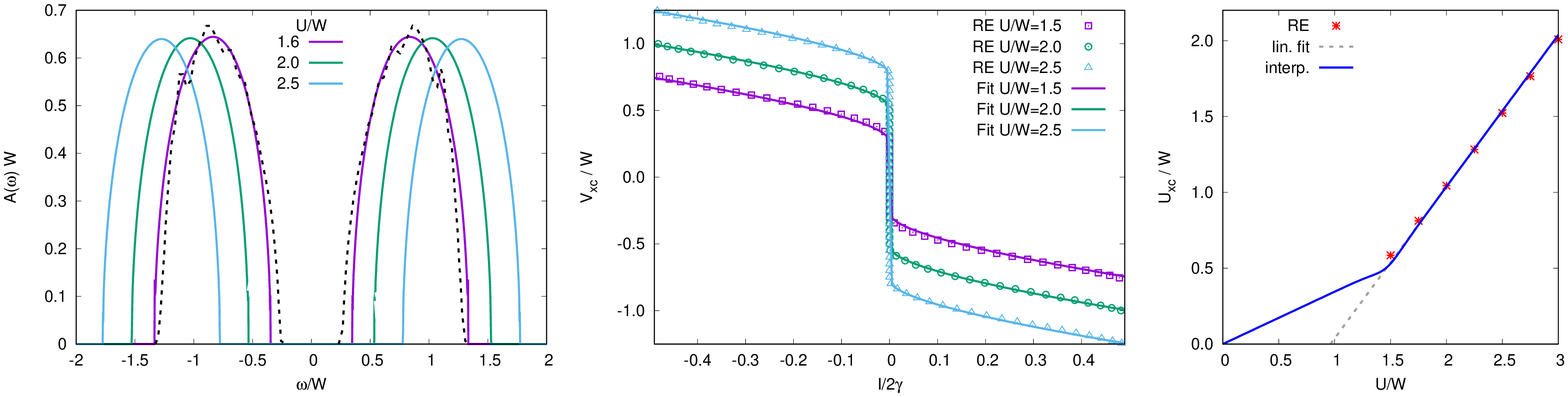}
  \caption{ \label{BL}
    Spectral functions and reverse-engineered xc-bias for the BL. 
    Left panel: Spectral functions calculated within DMFT+NCA for three values of $U$ (full colored lines)
    and within DMFT+NRG for $U/W=1.6$ (black dashed line, data taken from Ref.~\cite{Zitko:PRB:2009a}).
    Middle panel: Reverse-engineered (RE) xc-bias $V_\xc[I]$ obtained from DMFT+NCA spectral functions
    given in left panel. Full color lines show the parameterization of the xc-bias Eq.~(11) in main 
    text together with (\ref{gfunc}). Right panel: $U_\xc$ as a function of $U$ obtained from fitting Eq.~(11)
    to RE data.
  }
\end{figure}

\begin{figure}
  \includegraphics[width=\linewidth]{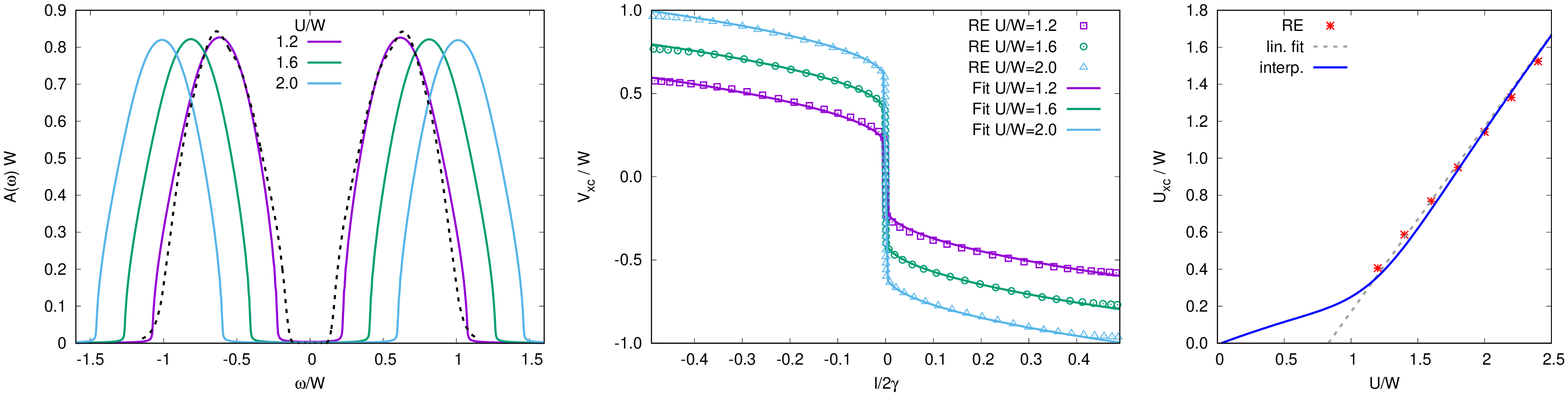}
  \caption{ \label{CL}
    Spectral functions and reverse-engineered xc-bias for the CL. 
    Left panel: Spectral functions calculated within DMFT+NCA for three values of $U$ (full colored lines)
    and within DMFT+NRG for $U/W=1.6$ (black dashed line, data taken from Ref.~\cite{Zitko:PRB:2009b}).
    Middle panel: Reverse-engineered (RE) xc-bias $V_\xc[I]$ obtained from DMFT+NCA spectral functions
    given in left panel. Full color lines show the parameterization of the xc-bias Eq.~(11) in main 
    text together with (\ref{gfunc}). Right panel: $U_\xc$ as a function of $U$ obtained from fitting Eq.~(11)
    to RE data.
  }
\end{figure}

In order to obtain the Mott part of the xc-bias functional $\bar{V}_\xc[I]$ (see Eqs.~(11,12) in the main text)
we have performed DMFT calculations in the Mott phase using the non-crossing approximation (NCA)~\cite{Pruschke:PRB:1993}
as impurity solver, which yields a reasonable description of the spectral function.
The left panels of Fig.~\ref{BL} and \ref{CL} show spectral functions for the Hubbard model on the Bethe lattice (BL)
and the cubic lattice (CL), respectively, calculated within DMFT+NCA for several values of $U$.
Also shown are reference spectral functions (black dashed lines) computed within DMFT+NRG by Zitko {\it et al.}~\cite{Zitko:PRB:2009a,Zitko:PRB:2009b}
which show overall good agreement with the corresponding DMFT+NCA spectra.

From the spectral functions computed within DMFT+NCA we can reverse-engineer the corresponding xc-bias
functional for computing the spectra in the ideal STM setup. To this end the STM tip is coupled to a 
single site of the Hubbard model. In this setup the Meir-Wingreen expression for computing the
current becomes
\begin{equation}
  \label{current}
  I = 2\gamma \int \frac{d\omega}{2\pi} \left[ f(\omega-V) - f(\omega) \right] A(\omega) = 2\gamma \int_0^V \frac{d\omega}{2\pi} A(\omega)
\end{equation}
where $f(\omega)$ is the Fermi function and $A(\omega)$ is the spectral function
of the site which in the ideal STM limit becomes the equilibrium one. In the last step we have assumed the zero temperature 
limit. The corresponding KS current is given by
\begin{equation}
  \label{ks-current}
  I_s = 2\gamma \int_0^{V_s} \frac{d\omega}{2\pi} A_s(\omega)
\end{equation}
where $A_s(\omega)$ is the KS spectral function and $V_s$ is the KS bias. 
Here the KS spectral function is either the semi-circular DOS of the infinite-dimensional BL or the one of the cubic lattice.

We can now obtain the KS bias $V_s$ and thus the xc bias $V_\xc$ from the many-body current 
as a function of the applied bias $V$ in the following way: For every bias $V$ we compute the many-body current
$I$ from Eq.~(\ref{current}). Now the KS bias $V_s$ is that bias for which the KS current $I_s$ exactly reproduces 
many-body current $I$, i.e. $I_s(V_s)=I(V)$. Hence numerical inversion using the bisection algorithm of the
expression for the KS current Eq.~(\ref{ks-current}) for $I_s=I$ yields the KS bias and thus the xc bias. 
The procedure can be summarized by the following scheme:
\begin{equation}
  V \ra I \equiv I_s \ra V_s \ra V_\xc=V_s-V
\end{equation}
The middle panel of Fig.~\ref{BL} shows the xc-bias obtained by this procedure from the many-body 
spectral functions of the Hubbard model on the BL for different values of $U$ in the Mott-insultaing regime.

\subsection{Parametrization of the xc-bias in the Mott phase}

We have found an accurate parametrization for the function $g$ of $\bar{V}_\xc$
(see Eq.~(11) of main text) for both the BL and the CL as 
\be
g(\tilde{I}) = 
\left((1-b)\sqrt{|\tilde{I}/2|}+b|\tilde{I}|\right)
{\rm sign}(\tilde{I})\,.
\label{gfunc}
\ee
where $b$ is an adjustable parameter between 0 and 1. For the BL we obtain excellent 
fits of the xc-bias in the insulating phase for $b=1/4$ (see middle panel of Fig.~\ref{BL}).
For the CL excellent fits (see middle panel of Fig.~\ref{CL}) are obtained for $b=0$ 
(such that the linear contribution to the function $g$ vanishes). 
Fitting our parametrization of $\bar{V}_\xc$ to the reverse-engineered xc-bias for different
values of $U$ then yields $U_\xc$ as a function of $U$ (see right panels in Figs.\ref{BL},\ref{CL}).
For both the BL and the CL $U_\xc(U)$ is approximately linear in the insulating (I) phase.
Linear fits yield $U_\xc^{\rm I}(U) = U - 0.96\,W$ for the BL and $U_\xc^{\rm I}(U) = U - 0.83\,W$ for the CL.

In the metallic (M) phase we cannot obtain the $U_\xc$ for the Mott part of our xc-bias functional 
easily since by definition the Mott gap vanishes. However, we have found that for our approach of
determining the Kondo prefactor from Fermi liquid conditions to work $U_\xc$ should not vanish at
finite $U$, but should vanish as $U$ approaches 0. This means that in the metallic regime $U_\xc(U)$ 
should not be interpreted as the xc discontinuity of DFT. Here we assume the following 
form in the metallic phase: $U_{\rm xc}^{\rm M}(U)=U_{\rm xc}^{\rm I}(U_c)\cdot(U/U_{c})$ which guarantees
continuity of the function at the critical $U$: $U_{\rm xc}^{\rm M}(U_c) = U_{\rm xc}^{\rm I}(U_c)$.
Finally, in order to obtain a smooth parametrization of $U_\xc(U)$ we interpolate between both 
regimes:
\be
U_{\rm xc}(U)=f(U-U_{c})U_{\rm xc}^{\rm M}(U)+[1-f(U-U_{c})]U_{\rm 
xc}^{\rm I}(U)
\label{uxc_bethe}
\ee
where $f(E)$ is the Fermi function at inverse temperature $\b=4/W$ and $U_c=1.47\,W$ for the BL
and $U_c=1.17\,W$ for the CL. The interpolated function $U_\xc(U)$ is shown in the right panels
of Figs.~\ref{BL} and \ref{CL} as blue lines.

\subsection{Quasiparticle weight for the cubic lattice}

The quasiparticle weight $Z(U)$ is an important ingredient of our parametrization of 
the xc-bias functional as it determines the Kondo prefactor parameter $\lambda_K$, c.f. Eq.~(9)
in the main text. For the BL the quasiparticle weight has been accurately calculated within DMFT+NRG
in Ref.~\cite{Bulla:PRL:99}. We find that $Z(U)$ is well fit by a shifted Lorentzian of width $\G_{z}=0.943\,W$:
\be
Z(U)=\left(1+\frac{\G_{z}^{2}}{U_{c}^{2}}\right)
\left(\frac{\G_{z}^{2}}{U^{2}+\G_{z}^{2}}-\frac{\G_{z}^{2}}{U_{c}^{2}+\G_{z}^{2}}
\right).
\ee
This is shown in Fig.~1 in the main text. 

\begin{figure}
  \includegraphics[width=\linewidth]{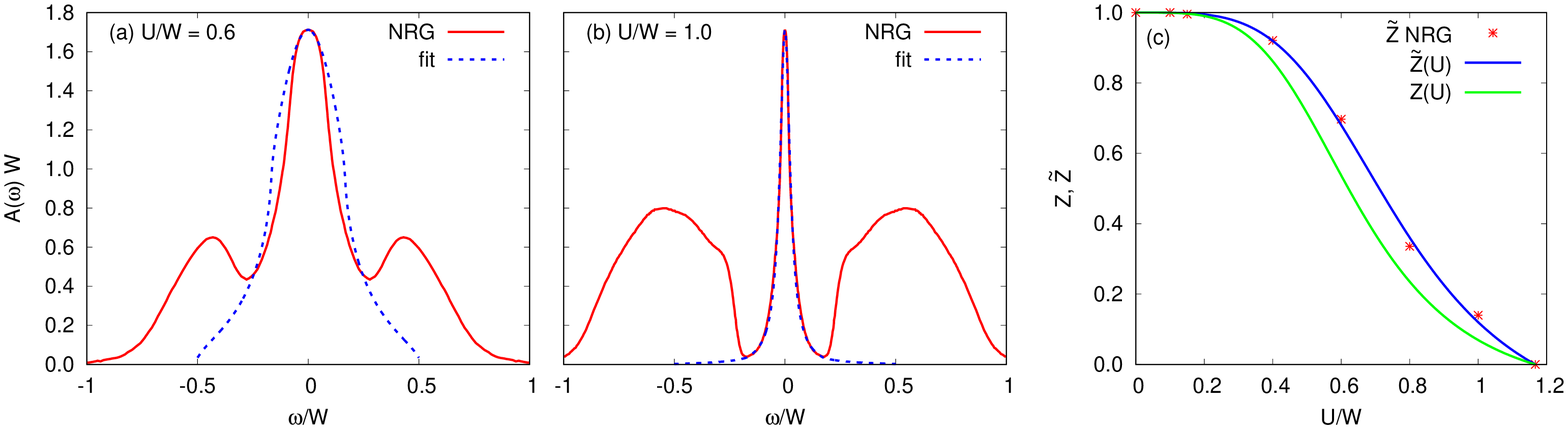}
  \caption{ \label{Ztilde}
    Pseudo quasiparticle weight for the cubic lattice. 
    (a,b) Fitting of quasiparticle peak according to (\ref{Aapprox}) to DMFT+NRG spectral function (taken from Ref.~\cite{Zitko:PRB:2009b})
    for two different values of the interaction $U$. (c) Pseudo quasiparticle weight $\tilde{Z}$ obtained from fitting (\ref{Aapprox}) for
    different values of $U$ in the metallic phase (red stars). The blue line shows a fit for $\tilde{Z}(U)$ assuming a ``quartic'' Lorentzian
    according to (\ref{Z-U_CL}) for $Z(U)$ (green line), related by (\ref{Ztilde-Z_CL}). The resulting fit parameters are: $\Gamma_z=0.65\,W$
    and $U_c=1.17\,W$.
  }
\end{figure}

For the CL, on the other hand, we do not have accurate data on $Z(U)$. However, since in Ref.~\cite{Zitko:PRB:2009b}
spectral functions were calculated within DMFT+NRG, we can instead directly obtain the pseudo 
quasiparticle weight $\tilde{Z}(U)$ by fitting the pseudo quasiparticle spectral function given by (\ref{Aapprox})
to the spectra close to $\omega=0$ whereby the non-interacting hybridization function (or embedding self-energy) is 
obtained from the non-interacting GF as
\be
\D_0(\w) = \L_0(\w) - i\G_0(\w)/2 = \w - [G_s(\w)]^{-1}
\ee
The latter can be computed analytically from the complete elliptic integral of the first kind, see e.g. Ref.~\cite{Zitko:PRB:2009b}.
Fig.~\ref{Ztilde} shows fitted pseudo quasiparticle spectra according to (\ref{Aapprox}) for two different values of $U$ (left and 
middle panel). Note that the fitting has been done such as to reproduce the curvature $A^{\prime\prime}(\omega)$ at $\w=0$. Thus for 
smaller values of $U$ the quasiparticle appears to have a considerably larger width than the actual spectral function.
The right panel of Fig.~\ref{Ztilde} shows the $\tilde{Z}$ obtained by fitting (\ref{Aapprox}) to the DMFT+NRG spectral functions
for different values of $U$.  

For the CL we have $G_0(0)\simeq-5.38i/W$ and $G_0^\prime(0)\simeq13.24/W^2$ and thus $G_0^\prime(0)/[G_0(0)]^2 \simeq -0.46$ which is very similar
to the corresponding value for the BL, leading to a similar relation between $\tilde{Z}$ and $Z$:
\begin{equation}
  \label{Ztilde-Z_CL}
  \tilde{Z}  \simeq \frac{2.2\cdot Z}{1.2\cdot Z +1}
\end{equation}
We next assume some function for $Z(U)$ and then fit the resulting $\tilde{Z}(U)$ according to (\ref{Ztilde-Z_CL}).
For the CL we find that the form
\begin{equation}
  \label{Z-U_CL}
  Z(U) = \left( 1+\frac{\Gamma_z^4}{U_c^4} \right) 
  \left( \frac{\Gamma_z^4}{U^4+\Gamma_z^4} - \frac{\Gamma_z^4}{U_c^4+\Gamma_z^4} \right)\,,
\end{equation}
leads to a very good interpolation of the numerically extracted values
of $\tilde{Z}(U)$ (in particular, it describes well the plateau-like behaviour
for small values of $U$). 

\bibliographystyle{prstyown}
\bibliography{supp}